\documentclass[aps,pre,twocolumn,showpacs,floatfix]{revtex4}

\usepackage{bm}
\usepackage{color}
\usepackage{amsmath}
\usepackage{graphicx}
\usepackage{epsfig}

\begin{document}

\title{The shower approach in the simulation of ion scattering from solids}

\author{V.A. Khodyrev$^1$}
\email{khodyrev@anna19.sinp.msu.ru}
\author{R. Andrzejewski$^2$}
\author{A. Rivera$^2$}
\altaffiliation{Present address: Instituto de Fusi\'on Nuclear, Universidad Polit\'ecnica 
de Madrid, E-28006 Madrid}
\author{D.O. Boerma$^2$}
\author{J.E. Prieto$^{2,3}$}
\email[Corresponding author. Email: ]{joseemilio.prieto@uam.es}

\affiliation{$^1$Institute of Nuclear Physics, Moscow State University, Moscow 119899, Russia \\
$^2$Centro de Microan\'alisis de Materiales, Universidad Aut\'onoma de Madrid\\
$^3$Dpto. de F\'\i{}sica de la Materia Condensada and
Instituto Universitario ``Nicol\'as Cabrera'', \\
Universidad Aut\'onoma de Madrid, E-28049 Madrid, Spain}

\begin{abstract}
An efficient approach for the simulation of ion scattering from
solids is proposed.
For every encountered atom, we take multiple samples of its thermal
displacements among those which result in scattering with high
probability to finally reach the detector.
As a result the detector is illuminated by intensive ``showers",
where each event of detection must be weighted according to the actual 
probability of the atom displacement.
The computational cost of such simulation is orders of magnitude lower
than in the direct approach and a comprehensive analysis of multiple 
and plural scattering effects becomes possible.
We use the new method for two purposes. First, the accuracy of the
approximate approaches, developed mainly for ion-beam structural
analysis, is verified.
Second, the possibility to reproduce a wide class
of experimental conditions is used to analyze some basic
features
of ion-solid collisions: the role of double violent collisions
in low-energy ion scattering; the origin of the ``surface peak" in
scattering from amorphous samples; the low-energy tail in the energy spectra
of scattered medium-energy ions due to plural scattering; the degradation of
blocking patterns in 2D angular distributions with increasing depth
of scattering. As an example of simulation for ions of MeV energies, we
verify the time-reversibility for channeling/blocking of 1 MeV
protons in a W crystal.
The possibilities of analysis that our approach offers may be very
useful for various applications in particular for structural
analysis with atomic resolution.
\end{abstract}


\maketitle

\section{Introduction}

The classical binary collision approximation is well applicable for 
the description of ion-solid interactions in the energy range above 
$\sim$1~keV; 
(in each ion-atom collision the interaction with nearby 
atoms is weak enough to allow to be treated as a perturbation if required).
This provides the possibility of an efficient reproduction of many 
experimental conditions playing with the parameters of the interaction model.
Due to the importance of this subject, big efforts have been directed
towards the development
of efficient simulation programs~\cite{Ekstein}. To illustrate the
central role that computer simulations have played in the field,
it will suffice to mention that one of the most prominent phenomena, the
channeling of ions in crystals, was first observed in simulation
results~\cite{RobinsonOen}. In this context, one should refer also to
the paper of Barrett~\cite{Barrett} which provides an important supplement
to the theory of ion-crystal interaction allowing to address some
aspects of the phenomenon which are difficult to treat theoretically.
Currently, with ion beams being widely used as a precision tool
for the analysis and modification of materials, there has been an increase
of the level of demands to simulation algorithms. One example is the
use of low-energy ion beams for surface structure determinations.
The classical picture of scattering together with the effects of blocking
of scattering on atomic pairs form the basis for obtaining detailed
information about the atom locations.
The only way to extract this information is by comparison of intensities
measured for different scattering geometries with the results of
simulations for many trial structures.

However, a fundamental problem occurs here due to the insufficient 
efficiency of existing algorithms. Due to the small scattering cross 
sections, the direct simulation by calculation of individual ion 
trajectories (the program MARLOWE~\cite{MARLOWE} is the most developed 
code of this type) is often not practicable.
To understand this one has to keep in mind that the experimental procedure
typically consists in the measurement of angular scans with a small-aperture
energy-resolving detector. To acquire the necessary statistics in a 
measured spectrum, the required number of ions in the beam amounts to 
$\sim 10^{9}$ for low-energy (keV range) scattering of heavy ions like 
Ne or Ar and to $\sim 10^{13}$ for scattering of ions with energies in 
the MeV range. It is clear that the direct simulation of such large 
number of ion trajectories is impossible, especially in the latter case.
For this reason, it is concluded (see, for example, 
Ref.~[\onlinecite{Ekstein}]) that even the power of supercomputers is by
far not enough to perform such simulations.

An overview of the existing approaches to this problem shows that
two main ideas are used depending on the ion energy.
In the simulation of backscattering of high-energy channeled ions,
one can rely on the single-scattering model assuming that
the motion of the ion in the outgoing path can be described by a
straight-line trajectory. This allows to avoid a precise description
of this segment of the ion trajectory. As a result, each ion contributes
to the statistics according to the probability of close collisions 
along the ingoing path.
The inverse (blocking) condition can be treated analogously
(see Ref.~[\onlinecite{Karamyan}], for example).
This model is well tractable and serves as the basis for a large number
of algorithms which have been developed (see
Refs.~[\onlinecite{FLUX,CRYSTALTRIM,UPIC}] for the most widely used).

This algorithm does not work, however, if one is interested in the simulation
of channeling-blocking conditions or in the scattering of low-energy or
heavy ions. To tackle this problem Tromp et al.~\cite{VEGAS} proposed to use
the property of ion motion known as the Lindhard time-reversibility
rule~\cite{Lindhard}. According to this property, the scattering yield can be 
obtained by a proper convolution of the flux of impinging ions with the 
time-reversed flux of ions imagined as being scattered into the detector 
aperture when these two fluxes meet in the sample volume.  In general, the use 
of this feature seems not to provide a possibility to facilitate the simulation: 
the convolution of fluxes in a six-dimensional phase space which, even at small
depths, can have complicated distributions, is also a hardly solvable problem. 
The procedure (further referred to as the ``reversing" procedure) is, however, 
heuristic in the sense that, in contrast with the direct simulation, it admits 
the use of certain approximations in the description of ion trajectories and 
this can be used to design much more efficient schemes of simulation.
In other words, one has here the possibility to dramatically boost the
efficiency, although this is achieved at the cost of partly sacrificing the 
accuracy of the description of the phenomenon, when this is admissible. 
In particular, a simplification of the description is applied in the algorithm 
of the program VEGAS~\cite{VEGAS}: the energy and angular variables 
are not considered in the flux convolution. This simplified procedure cannot 
reproduce all the details of multiple and plural scattering, as well as some 
specific features of energy losses. Fortunately, this turns out to be not a 
serious obstacle for the use of this program in the important application of
structure analysis using medium-energy ($\approx$100 keV) light ion 
scattering, MEIS.

On the other hand, the full version of the reversing approach is feasible
for the simulation of scattering of low-energy heavy ions (LEIS).
In this case, only a few layers at the surface of the sample contribute to the
scattering yield (this feature provides sensitivity to the surface structure)
and the structure of fluxes is not strongly developed. The latter circumstance
permits to use a coarse-grained representation of the fluxes and, as a result, 
their convolution becomes a tractable problem. Such full version of the reversing 
procedure is implemented in the program MATCH~\cite{MATCH} where
the convolution of fluxes is performed using a specific method: for two
sets of pre-calculated ingoing and outgoing trajectories, those pairs are
matched which can be connected as a result of the scattering on one target atom.

All the indirect methods of simulation mentioned above have rather restricted
regions of applicability. As a result, for many conditions where the multiple 
and plural scattering effects are significant, simulation results are now not
available. The development of efficient methods of simulation, if possible, 
is important for basic studies of ion-solid interaction and for the improvement 
of methods of analysis of materials based on the use of ion beams.

In the present paper we describe a new simulation method which use
the advanced possibility of the Monte Carlo method, the strategy of
importance sampling. This strategy is used in sampling of thermal
displacements of atoms met on the ion path.
This approach provides a possibility to increase the
efficiency of the direct method by several orders without necessity
to sacrifice the exact treatment of the binary collisions. 
The next Sec.~II describes how the importance sampling strategy can
be implemented in the simulation.
We give the name TRIC (Transport of Ions in Crystals) to the developed
computer program. 
In all other respects, we employ the standard
version of the binary collision model as used in the code MARLOWE.
Therefore, for all features of this model readers are referred
to Ref.~[\onlinecite{MARLOWE}]. In Sec.~III some results of the application
of the program TRIC are presented to demonstrate the efficiency of the
developed approach for the simulation of different experimental conditions.
In Sect.~IV we discuss the advantages of our approach over the simulation
algorithms proposed hitherto. Some conclusions are formulated in Sec.~V.

\section{The simulation procedure}

The ion velocity is assumed to be much higher than the thermal velocities of
crystal atoms, a condition which is usually well satisfied. This means that
the configuration of thermal displacements of all atoms can be predetermined
before the simulation of scattering of an individual ion. The configurations
must be chosen randomly according to the statistics of thermal vibrations.
However, it is a more convenient procedure, and therefore it is commonly used,
to choose the displacement of an atom met along the ion trajectory just
before the treatment of the collision. 
Notice that this procedure must take into account the correlation of thermal
displacements of different atoms as it is present in the lattice dynamics.
Notice also that, for any atom met by the ion, not only a single, but
also multiple samples of its displacement can be taken resulting in
different trajectories after the collision.
It can be argued that the simulation will correctly represent the statistics
of atom thermal vibrations if the contribution of each ion in the Monte-Carlo
sum is determined as the average of the results obtained in such multiple
trials~\cite{LTE}. This simple conclusion plays an important role in the
following consideration.

The direct simulation of histories of collisions with the actual distribution
of atom displacements is very inefficient because the number of successful
scattering events in all trials is minute (as in a real beam irradiation 
experiment). The main idea that allows to increase the efficiency is a 
separate treatment of those displacements that result in scattering 
events of the ion in the direction of the detector, because this fraction 
of ion trajectories has a much higher probability to actually end up there.
Moreover, by multiple sampling of such displacements for each atom the 
scattering flux can be increased even more.

\begin{figure}
\includegraphics*[width=140pt, angle=-90]{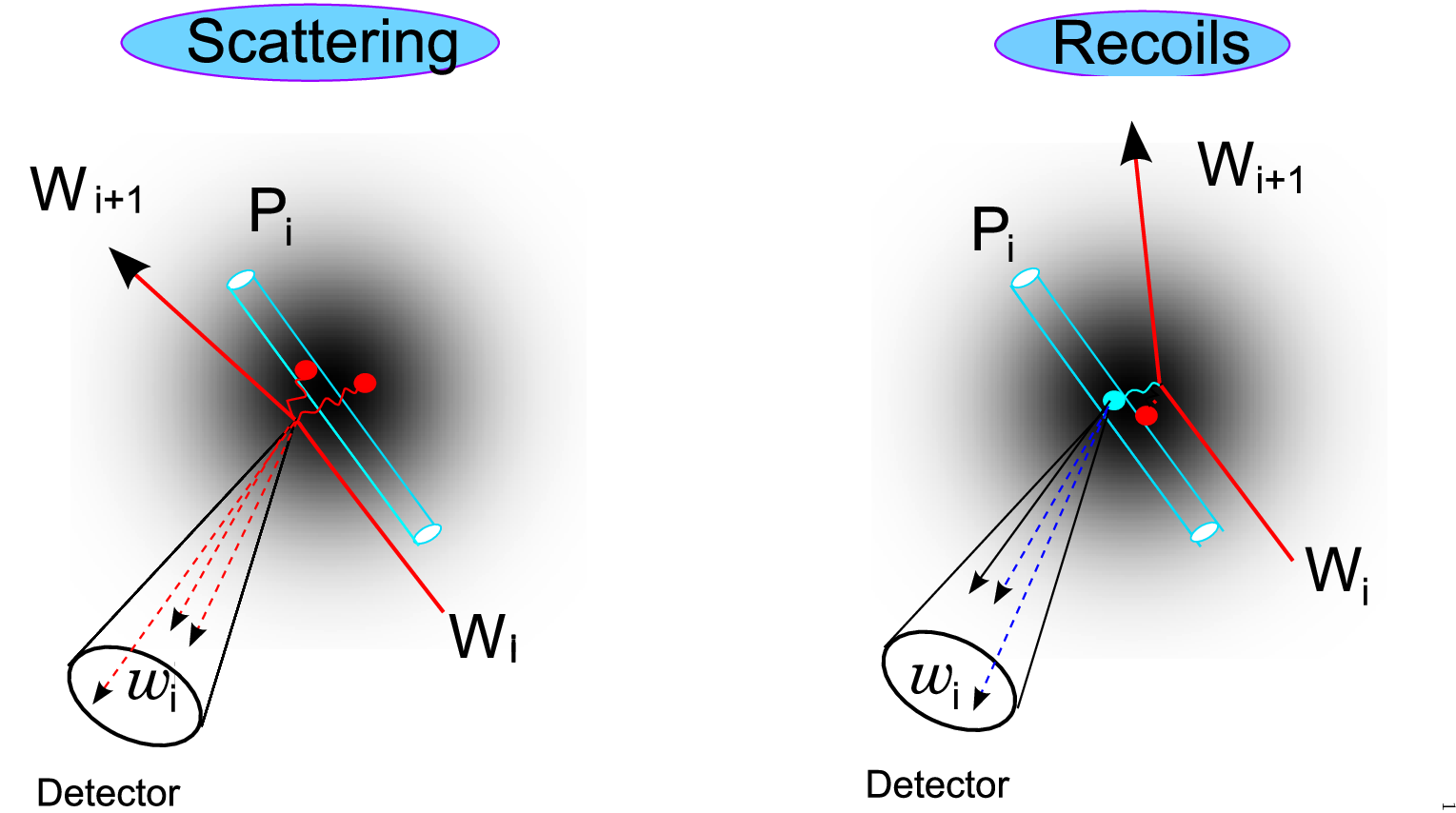}
\caption{\label{Fig1}
{\small (color online)
Illustration of the principle of selection of the ``hot" region of atomic
displacements
used in the shower generation. The density of the Gaussian distribution
of atomic displacements is represented by the gray cloud while the ``hot"
region is indicated as the tube enclosing the region of impact parameters
for scattering into the chosen angular cone (left panel).
Analogously, the picture at the right panel illustrates the generation
of showers of recoiled atoms. $W_i$ and $w_i$ are the weights ascribed
to the trajectories of primary and secondary ions, respectively, while
$P_i$ designates the integral probability of an atom displacement into
the ``hot" region (see the explanations following Eq.(2))}
}
\end{figure}

To implement this simple idea we use the following procedure. First, the
impact parameter $b_i$ which corresponds to scattering into the direction 
to the center of the detector (see the left-hand side of Fig.~\ref{Fig1}) 
is determined ($i$ denotes the number of the current lattice site).
Then, assuming that the interesting scattering directions lie within a 
cone of width $\Delta\Theta$, we can define the corresponding region of 
impact parameters whose half-widths in the scattering plane, $\Delta b_\|$, 
and in the perpendicular direction, $\Delta b_\bot$, can be estimated as
\begin{equation}\label{DDTT}
\Delta b_\|\approx|d\Theta/db|^{-1}\Delta\Theta,\,\,\,\,\,\,\,
\Delta b_\bot\approx (b/\sin\Theta)\,\Delta\Theta.
\end{equation}
(variation of $b_\bot$ means actually a rotation of the
scattering plane by an angle $\Delta \varphi\approx\Delta b/b$ with the
corresponding variation of the scattering angle
$\Delta\Theta=\sin\Theta\Delta\varphi$).
These conditions separate a region close to the ion trajectory, a tube aligned
with its velocity vector, and the associated fraction of the (Gaussian)
distribution of atom displacements.
Then, one or several ($n$) atom positions can be drawn according to the 
distribution with such cut-off. As a result, a ``shower" of ion trajectories 
is directed towards the detector. The remaining part of the distribution 
is used to continue the trajectory of the ``primary" ion. In such a way all 
possible displacements of the atom are sampled. Notice that the width of 
the shower $\Delta\Theta$ is assumed to be significantly larger then the 
width of the detector $\delta\Theta$.

In the accumulated statistics, events due to ions in the showers 
(``secondary" ions) must contribute with weights $w_i$ given by the 
probability for the atom to be displaced into the ``hot" region.
To calculate the weights we notice first that a weight should be 
ascribed also to the primary ion itself ($W_i$ before the $i$-th collision).
In every collision, this weight is decremented by the already considered
probability to scatter into the shower cone (trajectory ``degradation").
Then the weights $w_i$ and $W_{i+1}$ after the collision are updated as
\begin{equation}\label{DDWW}
w_i=P_iW_i/n,\,\,\,\,\,\,\,W_{i+1}=(1-P_i)W_i,
\end{equation}
where $P_i$ is the integral probability of atom displacement into the
``hot" region.  As it should be, the sum of probabilities is conserved: 
$W_{i+1}+nw_i=W_i$. It is worthwhile to note also that this approach 
provides the correct absolute value of the scattering yield: the 
expectation value for a certain energy-angular range is equal to that 
obtained in a direct simulation in which the same number of ions is sent 
to the sample.

A similar procedure can be used for the simulation of the yield of 
recoiled atoms. When the ion crosses a lattice site occupied by an atom of 
the considered species, a ``shower" of recoiled atoms is emitted. 
The ``hot" 
region (see the right-hand side of Fig.~\ref{Fig1}) selects now those 
atom displacements which result in emission of recoiled atoms within 
the chosen cone. To account for all possibilities of recoiling, one 
additional recoil atom is emitted by sampling the remaining part of the 
Gaussian distribution. The resulting recoil is allowed to produce new 
recoils, analogously as the ion itself, and also scattering showers.
Then, the trajectory of the ion is followed further with the
atom displacement drawn according to the total Gaussian distribution.
To describe the whole cascade, we consider also recoils produced by the
recoiled atoms in showers. The result of these many possibilities is a 
strongly developed tree of cascade. Since all particles in the cascades 
have similar histories of collisions, the treatment using a recursive 
algorithm turns out to be very efficient.

One can easily recognize the similarity of this approach with the
strategy of ``importance sampling" used in the Monte Carlo numerical
integration~\cite{MonteCarlo} where the values of the integration variable
are sampled more densely in the regions that give the highest contribution 
to the integral. This similarity is not accidental because the scattering 
yield, as an average over the ion initial conditions and the thermal 
displacement of atoms, is, in fact, determined by an integral over a 
multidimensional space. Due to the inherent complexity of this integral, 
the importance-sampling strategy can be used only on an intuitive basis 
as described above. In fact, we assume here that, even though the 
subsequent re-scattering of ions in the showers tends to diminish their 
intensities, the final effect will be a dramatic increase of the number 
of "detection events". It is clear that this strategy should be effective 
at high energies where the re-scattering events are of little importance. 
However, the shower approach turns out to be efficient also at low 
energies where re-scattering strongly modifies the flux of outgoing ions.
The reason is that, fortunately, this adverse circumstance is well
compensated by the fact that the primary ions themselves can leave with large
probability the sample volume and the numerous accompanying showers from
the top layers will produce again an intensive flux in the direction of 
the detector. We can say that, in this case, the showers are used to improve
the direct simulation at its last stage. These arguments show that the 
most serious difficulties in the application of the shower approach can 
be met in simulations for intermediate ion energies where the rare 
plural-scattering events result in violent fluctuations (see below how 
this problem can be eliminated).

Violent fluctuations in the accumulation of scattering events can have
different causes. A first circumstance which should be noted in this respect
is that the detection of the primary ion itself (with large weight $W_i$)
is completely excluded if the showers are generated at every lattice site
met along the ion path.
Thus, the displayed fluctuations are entirely due to the
dispersion of the values of $w_i$.

If the number of ions in one shower $n$ is fixed, the value of the weight
$w_i$ for a certain scattering angle $\Theta$ depends on the position of
the ``hot" region with respect to the center of the Gaussian distribution 
of thermal displacements. Since the width of this distribution is rather 
small compared to inter-atomic distances, variations of several orders 
of magnitude in the values of $w_i$ are possible. To improve upon this 
situation, it is reasonable, instead of fixing $n$, to fix the value 
of the weights, $w_i=w_0$, treating instead 
the number of ions in a shower as variable, $n_i=  P_i W_i / w_0$ (the 
fractional part is treated as the probability for sending one additional ion).
A reasonable value for $w_0$ can be estimated by defining a maximum value
for the number of ions $n_i$ in the most intense shower.
When $w_i\gg w_0$, the resulting effect is expressed as an increase of the
number of detected events with a smaller weight $w_0$ from intense showers
instead of one event with a large weight. An advantage in the inverse 
case, $w_i\ll w_0$, is that the load on the computer due to the simulation 
of non-significant events is avoided.

An additional advantage of the above approach is that the discrete counts of
such ``quanta of probability" closely mimics the aspect of experimental data.
The fluctuations of these counts are also the same as in the experiment.
Basically, the simulation performs two types of averaging: over the initial
conditions of impinging ions (the diversity of unperturbed trajectories of
the ions) and over the thermal displacement of target atoms which perturb
the ion motion. By variation of the value of the weight $w_0$, and
consequently, the number $n$ of ions in the showers, we can separately control
the fluctuations due to the finite statistics of atom displacements.

The fluctuations due to the finite number of impinging ions can be controlled
independently. The simplest way for doing this is just to increase the number
of ions (with the associated proportional enhancement of the computation 
effort). However, in the case of simulation for a sample with crystalline 
structure, a more efficient approach can be proposed where the same strategy 
of ``importance sampling" is applied. The contribution to the 
scattering yield from ions entering at different points of the crystal 
surface can be widely different. Therefore, it would be decisive to 
distribute the initial coordinates at the surface non uniformly, giving 
preference to those which result in an increased scattering yield.
Additionally, we must modify the initial values of the weight $W_0$ taking 
them as inversely proportional to the density of the distribution.
This can be organized as a self-adapting procedure; such an approach
is implemented in our computer code.

Another source of fluctuations is plural scattering. To illustrate
how these fluctuations arise, we should mention the following.
A reasonable criterion for the choice of the width of the showers
$\Delta\Theta$ is that the showers should be wide enough to encompass
the whole profile of multiple scattering on the outgoing path.
One may believe in this case that the transport into and the transport
out of the central region (of width $\delta\Theta$) are properly balanced
in the showers. However, at low energies and/or for heavy ions, this 
condition is difficult to fulfill because the showers should be
taken very wide and, as a result, only a small fraction of the ions in the
showers reaches our small-aperture detector. But, if the above condition
is not fulfilled, the reduction of ion flux in the shower cone due
to diffusion outward is to be compensated by the plurally scattered
primary ions. As this takes place, the main contribution comes from
primary ions scattered accidentally into directions close to the cone
mantle. Since the probability of further scattering into the cone
is large for these ions (small $|d\Theta/db|$ in (\ref{DDTT})) they
produce rare but very intense showers of secondary ions with almost
equal energies. As a result the accumulated energy spectrum is
disturbed by splashes in some channels.

We solve this problem in a similar way as described above: the
plural scattering events are sampled more frequently than they
happen in reality. For this goal, a second cone is considered,
coaxial with the first and significantly wider.  Wide showers are 
produced by the primary ion inside this cone in addition to the showers
in the internal cone. Then, ions of the outer cone are allowed
to send showers into the internal cone, similarly to the primary ion.
As a result, our goal is achieved because the weights of ions in the
outer cone are small compared with the weight of the primary ion $W_i$ and
the number of showers sent to the internal cone is now large enough to
sufficiently smooth the accumulated energy spectrum. In principle,
a whole hierarchy of such nested cones or even some smooth deformation
of the density of sampling of atom displacements could be organized 
(the strategy of stratified sampling~\cite{STRAT}). It turns out, however, 
that in many cases the implementation of the above-mentioned two-step 
approach solves practically all problems. This is illustrated in the 
next section (Fig.~\ref{Fig7}a).

The idea of stratified sampling has already been used in the simulation
of ion scattering~\cite{Jakas} though in a different form. The stratification
was applied as an adaptive procedure in the sampling of random numbers. 
The author used this method for the simulation of scattering from amorphous
samples using a Poissonian distribution for the 
inter-collision distances. The application of the strategy of stratified 
sampling as described above relies on physical arguments and, therefore, 
results in an efficient procedure which can be applied in general.

The last point that should be noted concerning the shower approach is that
the emission of showers aiming to the detector mainly results in misses.
This unavoidable drawback, when one is interested only in the yield on 
a small-aperture detector, turns out to be an advantage if one needs to 
simulate the 2D angular distribution of the scattered ions. Such possibility 
is illustrated in the following section. In fact, we are capable to calculate 
with the present approach 3D-distributions including the energy scale.

The implementation of this algorithm as a FORTRAN program incorporates all
important elements of the binary collision model~\cite{Ekstein}. A wide
possibility to choose the type of the interaction potential is provided. The
scattering integrals for binary collisions as functions of impact parameter
and energy are tabulated at a preliminary stage of the calculation and
used afterwards by interpolating with splines. Additionally, the 
impact-parameter dependence of inelastic energy losses is considered.
To account for the simultaneous interaction with two or more atoms,
we sum the deviations of ion motion due to the interaction with individual
atoms. In principle, all this provides the possibility to perform simulations
for energies in a wide range. 

At high energies the procedure of generation 
of showers can encounter difficulties due to restrictions in the accuracy of
computer calculations. The reason is that the width of the ``hot" region
in the transverse plane becomes so small that an accurate transformation of
impact parameter to scattering angle may be practically impossible.
For this reason, if this width becomes smaller than $0.01$ of the thermal
vibration amplitude $u_1$, the procedure of shower generation is
inverted: the scattering angle is sampled within the shower cone and then,
in reverse order, the impact parameter is calculated.

Materials of any crystal structure, including compounds, with rectangular
unit cells can be described in the input. On a lower level, however, the
program works with a description in terms of Wigner-Seitz cells.
In principle, this provides the possibility to treat also wider classes 
of structures. Amorphous structures can be simulated by a random 
rotation of the crystal lattice after each collision. The surface of 
the sample is defined as an imaginary plane appropriately located with 
respect to the crystal lattice. If the detector position is defined to 
be at the back side of the sample, a simulation of transmission through 
a crystal slab of a given thickness is performed. As output data, depth 
profiles of close collisions, energy spectra and 2D angular distributions
of scattered ions or recoiled atoms of a certain species are calculated.
Finally, the procedure is automated to calculate angular scans for a
step-wise rotation of the target, the detector or the beam direction 
around a given axis. The FORTRAN code of the program TRIC is supplied 
with a graphical user interface allowing to comfortably supply the 
input data, to run the calculations and to inspect the simulation results.
The software is available in Internet and all technical
details together with the description of the underlying physical model are
described in the supplied instructions.

The efficiency of the proposed approach depends on the
choice of the type of stratification, the width of the shower cones
and the numbers of ions emitted in each shower. 
It is hardly possible to give a ``universal" recipe 
for the choice of these parameters for a given condition. Our experience 
gained by the use of the program shows that, by estimating the possible 
role of plural and multiple scattering in the considered conditions, one 
can easily guess values for the parameters which are close to optimal.

As a measure of efficiency the flux directed to the detector should 
be estimated.
In the case wherein the flux of incoming ions is assumed to be uniform, the
yield of scattering from one atomic layer is easy to estimate. For example,
the yield of scattering of 100 keV protons from silicon within a cone of
10$^\circ$ width amounts to $\sim$10$^{-8}$ per incoming ion. 
Precisely the same yield
is reproduced in the direct simulation. On the other hand, it is well
realistic for the shower approach to obtain, on average, one ion
from each layer within a shower of the same width. 
Of coarse, the volume of calculation in this case is larger;
the same computations, together with the calculation of the trajectory of
the incoming ion, need to be performed for each ion in the showers.
For lower energies the relative efficiency decreases. For example, the 
yield of scattering of 10 keV Ar ions from one layer of copper is 
$\sim$10$^{-3}$.
This gain in efficiency is less expressive, however the problem
itself is also much easier. The increase of efficiency in the treatment
of the plural scattering is more significant: in the double-cone approach
the procedure of shower generation is performed twice. Applying the
stratification of initial conditions in the simulation of channeling we 
artificially increase the ion flux near the atom locations. As our experience 
shows, this can result in an additional increase of efficiency by one order 
of magnitude.

In general, the shower approach solves the main problem; simulations of
ion-solid collisions become possible, even using an ordinary PC.
In this way, all the results shown in the next section were obtained. The
required CPU times ranged from several minutes, as for the simulations of
low-energy ion scattering, to several hours, as for the data
shown in Fig.~8. In cases as in this example, wherein whole histories of
binary collisions need to be calculated up to large depths, the simulation
becomes a rather hard problem.

\section{Program tests and applications}

In this section we present several examples of applications of our program 
to simulations of different experimental conditions. The calculations 
were performed using simple and commonly used models of ion-atom interaction. 
At low energies, as in the cases shown in Figs. 2-4, the angle of deflection
$\Theta$ and the elastic energy loss in each collision were calculated using 
the ZBL potential~\cite{Universal}.
For determining the inelastic energy loss as a function of the impact 
parameter $\Delta E(b)$, the Oen-Robinson model \cite{RobinsonOen} was applied.
At higher energies, the Moliere potential was used and $\Delta E(b)$ was 
taken to be proportional to the electron density along the ion trajectory. 
In order to check the sensitivity of our results to the shape of $\Delta E(b)$,
we varied the screening parameter $a$ in the potential while determining the
electron density from the Poisson equation. For the type of data we simulated,
we found a negligible sensitivity and we chose for $a$ a value twice higher 
than its standard value. In all cases $\Delta E(b)$ was normalized to the 
Ziegler stopping cross section \cite{Universal}. To account for the energy 
loss straggling, we chose in each collision the value for the actual energy 
loss $T$ randomly according to the distribution for free electrons: 
$dP(T)=k/T^2$ ($T_{min}<T<2mv^2$), where $k=2\pi Z_1^2e^4 n/mv^2$, $Z_1$ 
and $v$ are the atomic number and velocity of the ion, respectively, and 
$n$ is the integral of the electron density along the ion trajectory. The 
minimal
value $T_{min}$ was determined by the condition that the mean value of $T$ 
must be equal to $\Delta E$. Such form of energy loss distribution has been
used previously (see Ref.~[\onlinecite{Tschalar}]). Finally, the multiple 
scattering on electrons was accounted for by a broadening of the deflection 
angle $\Theta$ with a gaussian distribution of width 
$\Delta \Theta=\sqrt{(m\Delta E)/(M_1E)}$. Here $M_1$ and $E$ are the mass 
and energy of the projectile, respectively.

\begin{figure}
\includegraphics*[width=8.5cm]{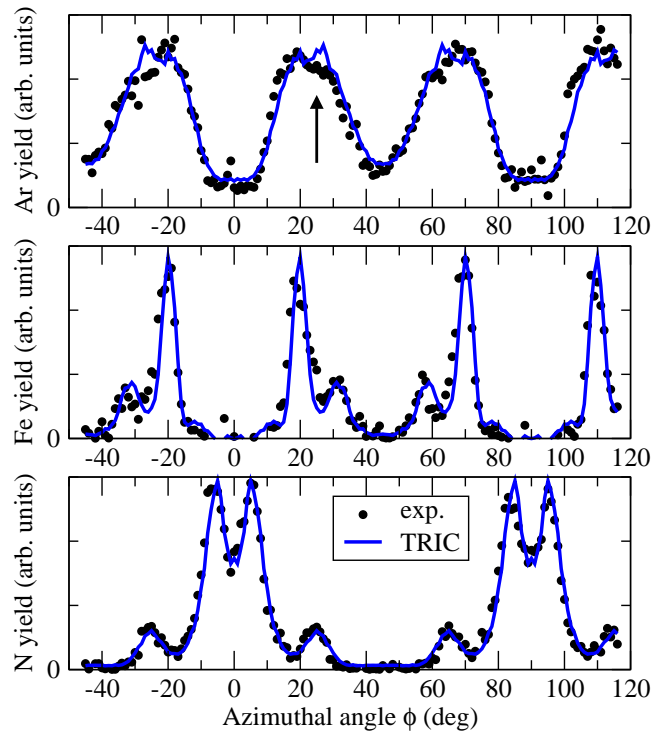}
\caption{\label{Fig2}
Comparison of simulated angular scans for the scattering of 5~keV Ar$^+$
ions from the surface of a Fe$_4$N(100) crystal with the experimental
results from Ref.[\onlinecite{Grachev}] (top panel). The yield of recoiled
Fe and N atoms is shown in the center and lower panels, respectively.
See text for more details.}
\end{figure}

\begin{figure}
\includegraphics[trim = -0 -10 0 0, scale = 0.8]{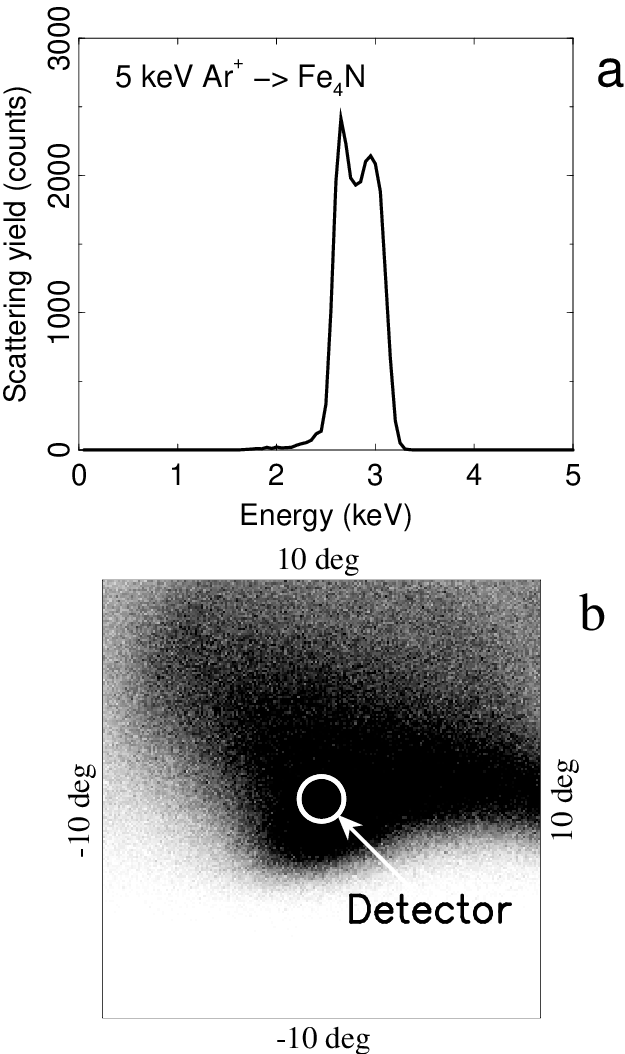}
\caption{\label{Fig3}
{\small
Energy spectrum of Ar$^+$ ions scattered from the (100) surface of a
Fe$_4$N crystal for the orientation indicated in Fig.~\ref{Fig2} by an arrow.
The lower part of the figure shows the angular distribution of scattered
ions. Results are shown using a gray scale where darker represents higher
intensity.
}
}
\end{figure}

The first example concerns the scattering of 5~keV Ar$^+$ ions from the (100)
surface of a Fe$_4$N crystal as studied experimentally in
Ref.~[\onlinecite{Grachev}]. The structure of this crystal can be considered
as {\em fcc}~Fe with an additional N atom located at the center of the unit
cell. The top layer of the stable (100) surface is the layer containing both
Fe and N atoms. Depending on the growth conditions, the surface can exist 
in two types of reconstructions, c4 and 4pg. Figure~\ref{Fig2} shows 
simulated angular scans for the c4 surface, which differs from the 
bulk-terminated one only by an outward displacement (of 0.23~\AA)
of the nitrogen atoms. In these scans, the crystal is rotated around
the surface normal, which is coplanar with the beam and the detector
directed respectively at 42$^\circ$ and 12$^\circ$ from the surface
at opposite sides of the normal. The yields of scattered Ar as well as of 
Fe and N recoils were calculated. In Fig.~\ref{Fig3}, the energy spectrum 
of scattered Ar is shown for the orientation indicated in Fig.~\ref{Fig2} 
by an arrow. Also shown is the angular distribution of scattered ions. 
The latter illustrates the blocking of scattering from the top-layer Fe 
atoms by re-scattering from nearby atoms; this largely explains the 
strong variation of the yield in the angular scan. These variations 
are strongest when either the beam or the detector are located close to 
the scattering ``horizon" which is formed by reflection from the surface 
as a whole (very small intensity at the bottom part of the angular 
distribution). In the case shown in Fig.~\ref{Fig3}, the detector is located 
at the edge of the shadow cone. The shape of the energy spectrum as an 
isolated peak is due to the blocking of the particles scattered from atoms 
of deep atomic layers by atoms of the top layer. This shape of the spectra 
favors the choice of the energy window for the performance of angular scans.
The double-humped structure is explained by the effective competition of 
single- and double-scattering from Fe atoms of the top layer~\cite{DoublePeack}.

As seen in Fig.~\ref{Fig2}, the experimental results are well reproduced 
in the simulation both for scattered ions and for recoils of the two 
atomic species. Note that, in order to achieve this agreement, the 
screening radii in the used ZBL potentials 
were properly reduced, as commonly done for the description of scattering 
of heavy ions at low energies. With the same potential, the data for the 
more complex reconstruction 4pg are also reproduced with the same 
quality. In general, the present results demonstrate that the quality
of the description of the experimental data is similar to that
provided by the code MATCH~\cite{Grachev}; in fact, the results of both
simulation can hardly be distinguished when plotted together.

\begin{figure}
\includegraphics*[width=8cm, angle=0]{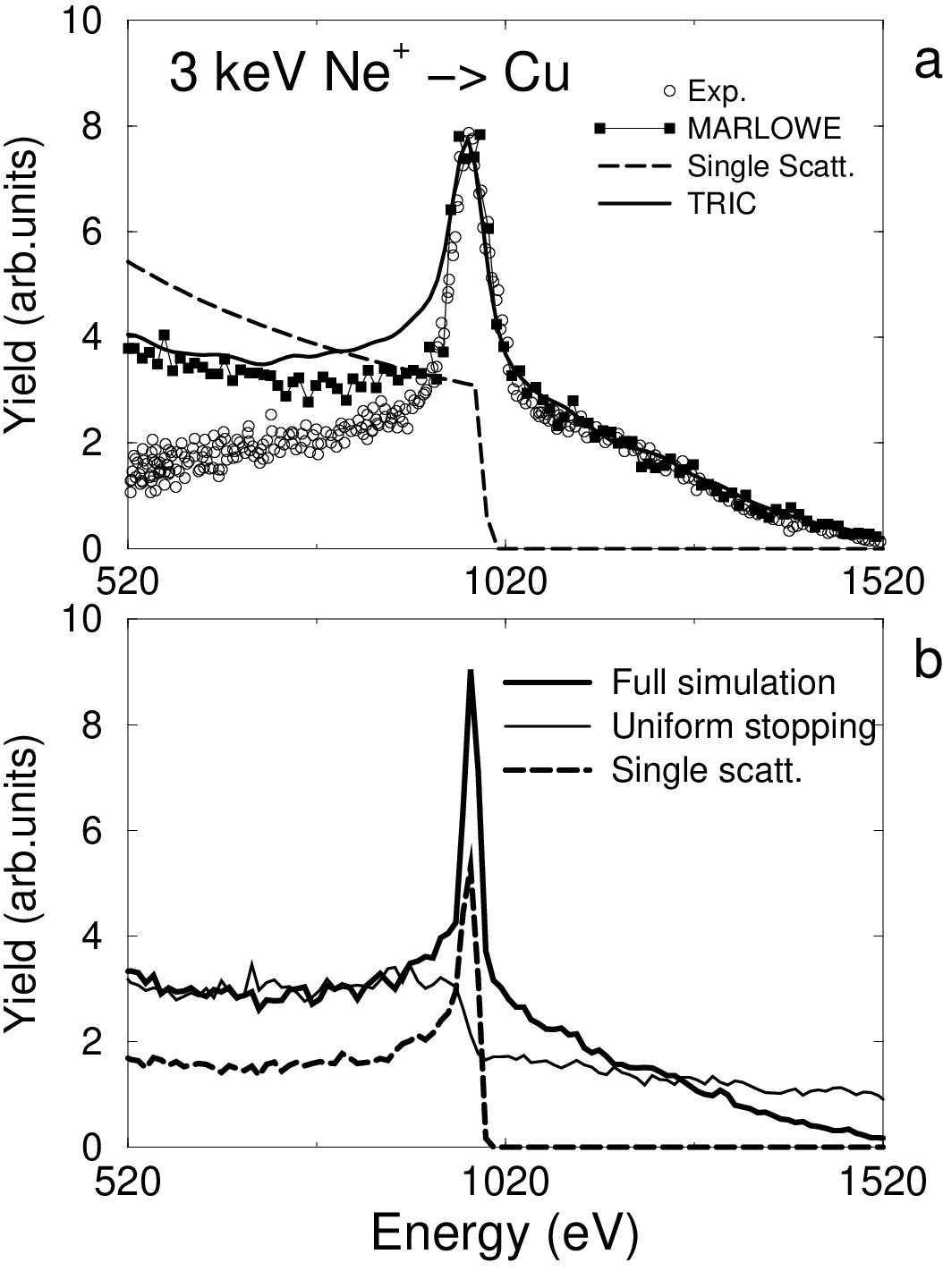}
\caption{\label{Fig4}
{\small
(a) Comparison of our simulation results for 3~keV Ne$^+$ ions scattering
from amorphous copper with experimental data~\cite{SurfPeak} and results
of simulation with the MARLOWE code.
The simulation results were folded with a gaussian distribution
in order to simulate the experimental energy resolution of 40~eV.
The dashed curve corresponds to a calculation in the single-scattering
approximation. Panel (b) shows simulations performed by applying
additional approximations.
The dashed curve corresponds to a simulation using the single scattering
approximation with impact-parameter-dependent energy losses while
the thin solid curve was calculated considering the energy losses as
the result of a uniform stopping force.}
}
\end{figure}

The second example, shown in Fig.~\ref{Fig4}a, corresponds also to a LEIS
experiment, now of 3~keV Ne$^+$ scattering from a polycrystalline Cu sample.
Under these conditions, an intriguing peak is observed~\cite{SurfPeak}
in the experimental spectrum at the energy corresponding to a single
Ne-Cu collision. Here the beam incidence was along the surface normal and
the scattering angle was 129$^\circ$. For these strong-scattering
conditions the code MARLOWE is capable to reproduce the general shape of 
the spectrum including the surface peak. Our simulations shown in the 
same picture also reproduce the spectrum shape; some differences with 
the MARLOWE simulation at low energies may be due to a difference in 
the potentials used for the description of the ion-atom interaction 
or in the treatment of inelastic energy losses. In general, the shape 
of the spectrum under these conditions differs considerably from that 
predicted by the single-scattering model (also shown in the figure).
The authors of Ref.~[\onlinecite{SurfPeak}] associated the appearance 
of the peak with the onset of plural and multiple scattering in the deeper 
layers. Here, using our advanced possibilities for simulation, we are able 
to come to more detailed conclusions about this striking phenomenon.

From general considerations one can conjecture the following two reasons for
the formation of the surface peak. First, due to the large variations of 
kinematic energy losses, the plural collisions yield a broad energy 
spectrum in scattering (in particular, this explains the presence of the 
high-energy tail), while ions which undergo only one close collision with 
atoms (single-scattering fraction) have a principally different energy 
distribution. It can be assumed in the considered case that, due to the 
strong scattering, only the ions coming from shallow depths can leave the 
sample without additional re-scattering. Thus all these ions have almost 
equal energies and form a peak in the energy spectrum which is superimposed
on the plural-scattering background. On the other hand, the surface peak 
can appear even in the single-scattering model if the impact-parameter 
dependence of the energy loss is taken into account. Indeed, in the case 
of sharp localization of energy loss at small impact parameters, ions 
scattered at shallow depths often do not experience significant energy 
losses on both ingoing and outgoing paths. So, again, many scattered ions 
have almost the same energies near the high-energy edge of the energy spectrum.

To establish the relative role of these two effects we performed simulations
with certain modifications of the model of interaction. Fig.~\ref{Fig4}b shows
the results of a simulation where both nuclear and electronic energy losses 
are replaced by equivalent continuous stopping forces (thin solid curve).
As seen, using such description of energy losses, the spectrum changes 
drastically. First, the surface peak disappears since ions 
scattered from different depths have now different energies. And second, 
the variation of kinematic energy loss in plural scattering is now not 
effective and, as a result, the shape of the spectrum on both sides of 
the surface peak is also strongly modified. On the other hand, in the 
shower approach we can also check the role of plural and multiple scattering.
For this purpose, a simulation was performed using a special
procedure, in which in all collisions before and after the emission of the 
shower, the ion deflection was canceled. The difference with the ordinary 
single-collision model is that here, in all collisions, the energy loss is
treated as a function of the impact parameter. These results are also shown 
in Fig.~\ref{Fig4}b (dashed curve) and demonstrate that the multiple and
plural scattering events themselves are of minor importance for the
formation of the surface peak.
In summary we can conclude that the surface peak reflects mainly the
correlation between scattering and energy loss: at these low energies, ions
scattered in deeper layers have more chances to exit from the sample if
they do not re-scatter strongly on atoms of the upper layers and,
consequently, the energy losses in the passage through these layers
are also small. At energies below the surface peak, both the present 
and MARLOW simulations differ significantly from the experimental results. 
A discussion of the difference is out of question, however, because the 
shape of the experimental spectrum was not reliably determined due to the 
uncertainty in the efficiency of the detector employed. On the other 
hand, there are also significant differences between the results of both 
type of simulations.
Assuming that both simulation codes perform a reliable treatment of the
problem, the difference must be ascribed to the use of different models
of ion-atom interaction (interaction potentials or inelastic energy losses).

It is worthwhile to note that the program TRIC passes here a serious test:
the simulation with shower generation produces results identical to those 
obtained when running the program in the direct simulation mode.
\begin{figure}
\includegraphics*[trim = -20 -20 -40 -20, scale = 0.6]{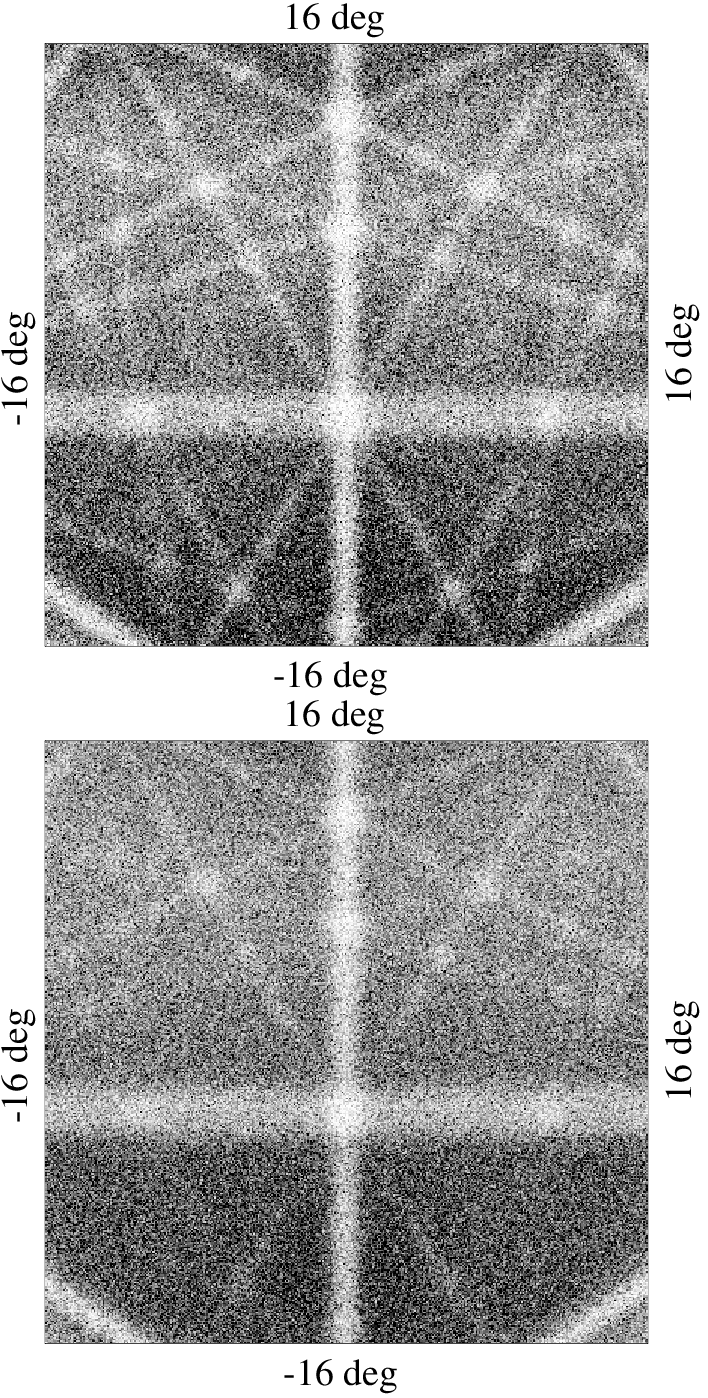}
\caption{\label{Fig5}
{\small
Results of the simulation of 100~keV He$^+$ ions scattering from a Si crystal.
The scattering yield is accumulated from depth ranges of (a) 5-30 nm
and (b) 30-55 nm. The displayed angular
range lies around the $<$112$>$ crystal direction.}
}
\end{figure}

Turning now to medium energy ion scattering (MEIS), we show in Fig.~\ref{Fig5}
the angular distribution of 100~keV He$^+$ ions scattered on a Si(100) 
crystal. The geometry (the $<$112$>$ axis is in the center of the 
position-sensitive detector) and the two depth ranges chosen for collection 
of the scattering events are equivalent to the experimental conditions 
used by Kobayashi~\cite{Kobayashi}. The results of the simulations are also
very similar to the experimental results: even at such relatively small
depths, the washing-out of the blocking pattern is well seen, first of all 
for the narrow channels. Although this effect of re-channeling is well 
predictable, its detailed demonstration in the referred experiment is 
rather interesting and our simulations support these results.
Notice that, with $10^5$ ions being sent to the crystal in the simulation,
the total yield over the detector amounts here to $\sim 0.01$ only
(remind that, in the direct simulation, this would be an expected number
of counts). In this sense, the results of our simulation are unique, it 
is hardly possible to obtain them using other known methods.
\begin{figure}
\includegraphics*[width=8cm, angle=0]{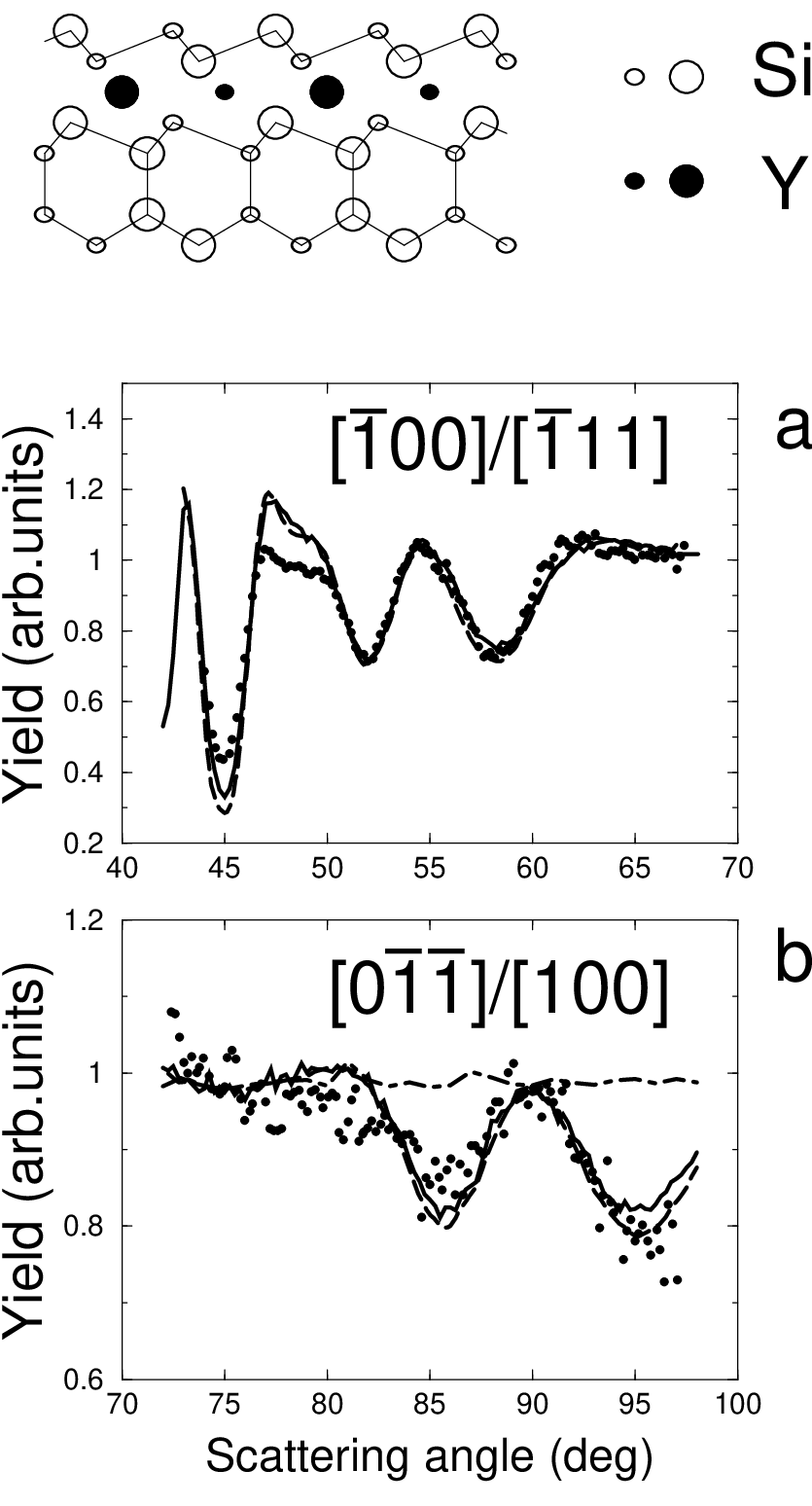}
\caption{\label{Fig6}
{\small
Comparison of our simulations (solid curves) of angular scans for 100~keV
$p$ scattering from a monolayer-thick YSi$_2$ film on Si(111) with the
result of a VEGAS simulation (dashed curves) and experimental data~\cite{Tear}
(dots).
The direction of the incident beam is parallel to (a) the $<$\={1}00$>$
and (b) $<$0\={1}\={1}$>$ directions of the Si substrate while the detector
lies in the plane $($01\={1}$)$.
The structure of surface layers is shown at the top. The yield is
normalized to the Rutherford cross section. The dash-dotted curve in the
lower graph shows results for a trial structure with Y atoms located at
the top layer.}
}
\end{figure}

In Fig.~\ref{Fig6} we show MEIS angular scans for the yield of 100~keV 
protons scattered from Y atoms of a YSi$_2$ monolayer epitaxially grown
on Si(111) (a side-view of the structure is shown in the picture).
The results of the measurements and of the simulations with the program 
VEGAS are taken from Ref.~[\onlinecite{Tear}]. The dips in the scans 
are due to blocking of the scattering from Y atoms by Si atoms of the 
upper layers. All parameters in the two simulations are taken to be 
identical. The achievement of a perfect agreement by optimization 
of the parameters of crystal structure and lattice dynamics is, in fact, 
the actual goal of the MEIS analysis~\cite{Tear}. Here, we emphasize 
only that results of our simulation coincide well with the results of 
the VEGAS simulation. In principle, this is predictable for such thin 
layers (see Introduction). As an additional test of our simulation 
procedure, we performed simulations also for the case of the top layer 
being terminated by Y atoms instead of Si atoms. As seen in the figure, 
the scattering yield in this case simply reproduces the Rutherford 
cross section.

\begin{figure}
\includegraphics*[scale = 0.5]{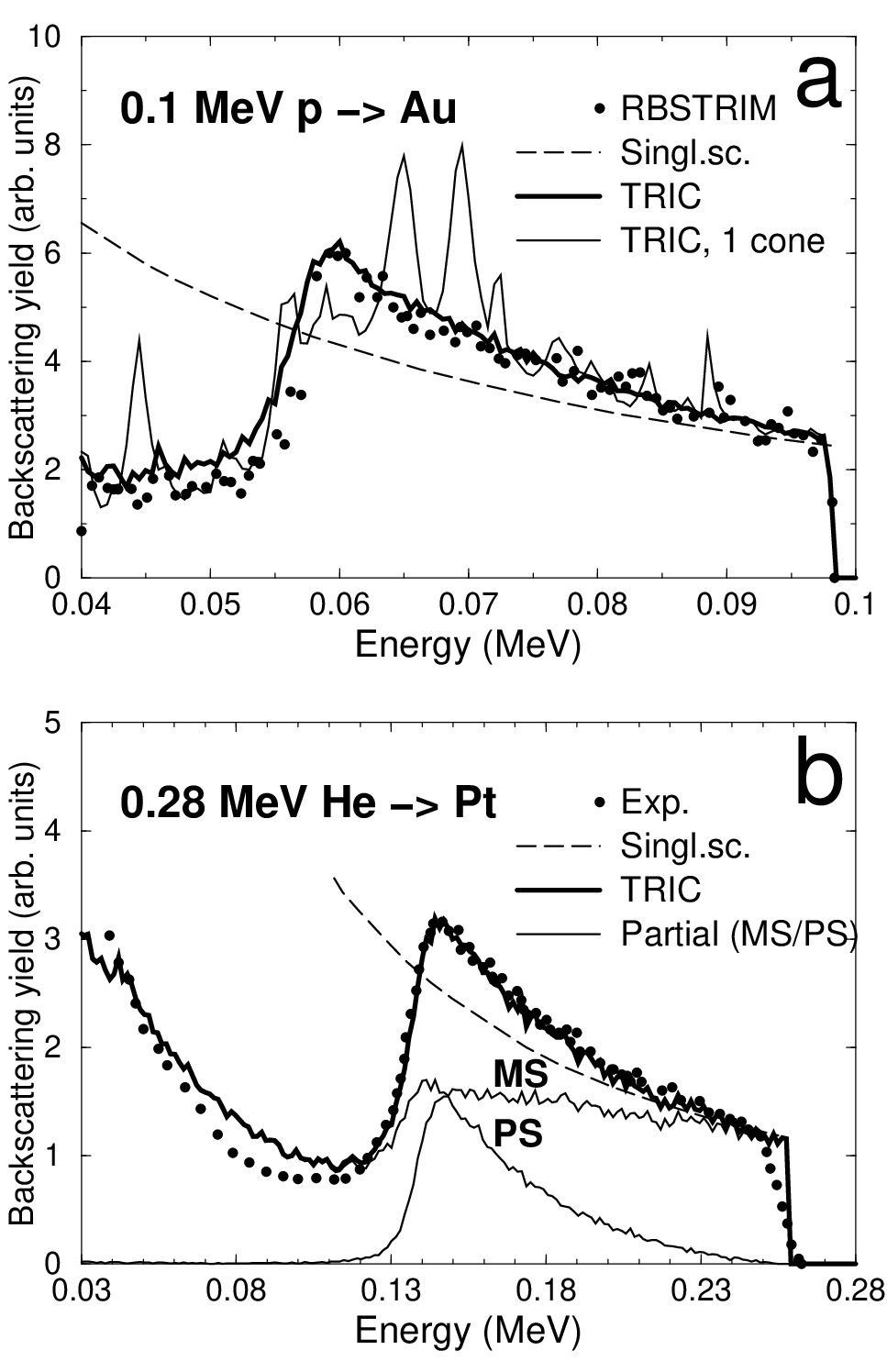}
\caption{\label{Fig7}
{\small
Backscattering energy spectra of 0.1~MeV~$p$ scattered from a 1000~{\AA } Au
film (a) and of 0.28~MeV He ion scattering from a 1130~{\AA } Pt foil (b).
The simulation results and the calculations in the single-scattering
approximation are compared with the RBSTRIM simulation~\cite{Biersack1}
and with experimental data~\cite{Chu}. Results of simulation in the
single-cone approximation are shown in the top panel. The lower graph
shows the partial contributions of ions from
the internal (MS) and outer (PS) cones.}
}
\end{figure}

The most difficult problem for simulations is the description of scattering in
conditions where the contribution from plural scattering is significant.
Besides low energies, this happens also in the medium and high-energy cases
where heavy-ion beams and/or heavy-atom targets are considered.
For amorphous samples, a well valid approach is to
consider the sample as a continuous media and to draw the path lengths
between close ion-atom collisions according to a Poisson distribution. 
Many simulation programs of such type have been developed (see 
Ref.~[\onlinecite{Rauhala}] for a review) which include different 
possibilities to treat multiple and plural scattering.
The most developed model of interaction is used in the widely known code
TRIM~\cite{Universal} but its efficiency is not sufficient for simulation
of backscattering.
Biersack {\it et al.}~\cite{Biersack} found a possibility to
increase the efficiency of this code up to an appropriate level without
a significant loss of accuracy of the description.
Figure~\ref{Fig7}a shows their results~\cite{Biersack1} for
scattering of 100~keV protons from a 1000~\AA-thick gold foil, together with
the results of simulations done with the shower approach. As seen, the
two approaches produce in fact identical results, also with similar
computation efforts. The only visible difference is at the low-energy tail
where the yield is entirely due to plural scattering. Therefore, these
results can be regarded as a confirmation of the accuracy of the TRIM
approach by comparison with an exact treatment of the model with a similar 
structure of the target.

Figure~\ref{Fig7}b shows the simulation results compared with results of an
experiment~\cite{Chu} of 280~keV $\alpha$-particles scattering from a
1130~\AA-thick Pt foil. In general, the agreement is satisfactory also 
here. In the two cases shown in Fig.~\ref{Fig7}, the shapes of the 
spectra differ significantly from those predicted by the single-collision 
model: the yield is higher for the main part of the spectra, also the 
low-energy tail, which is entirely due to the plural scattering, cannot 
be predicted at all by the single-collision model. To reproduce all 
these features with an accuracy above the fluctuation level, the showers 
must be generated in wide cones.
In the cases considered, the double-cone approach (see Sect.~II) was used
with a half-width of 40$^\circ$ for the internal cone and of 160$^\circ$
for the outer cone. In this way, at least the double-scattering is
treated with special efforts. The effect of the use of the two cones is
illustrated in Fig.~\ref{Fig7}a where, for comparison, the simulation results
obtained in the single-cone approach are included. The level of fluctuations
in this case suggests that much larger efforts are necessary to achieve the
same level of precision of simulation as in the double-cone approach. To
demonstrate the role of plural scattering in more detail, 
we show (Fig.~\ref{Fig7}b) separately the contribution of showers emitted 
by ions moving in the outer cone. It is seen that these histories of 
collisions completely explain the nature of the low-energy tail and, 
in general, contain mainly the effect of plural scattering. On the other 
hand, the partial spectrum of ions in the showers produced directly by 
the primary ion is influenced by multiple scattering. It looks surprising 
at a first glance that, in contrast with the predicted increase of the 
yield compared to the single-scattering spectrum~\cite{KhoMult,Smith}, 
this partial spectrum shows the inverse ordering due to multiple scattering. 
This is, however, natural because in this partial spectrum we do not 
consider the compensation of the transport out of the internal cone by
the counter-transport from the external region.
\begin{figure}
\includegraphics*[width=200pt, angle=-90]{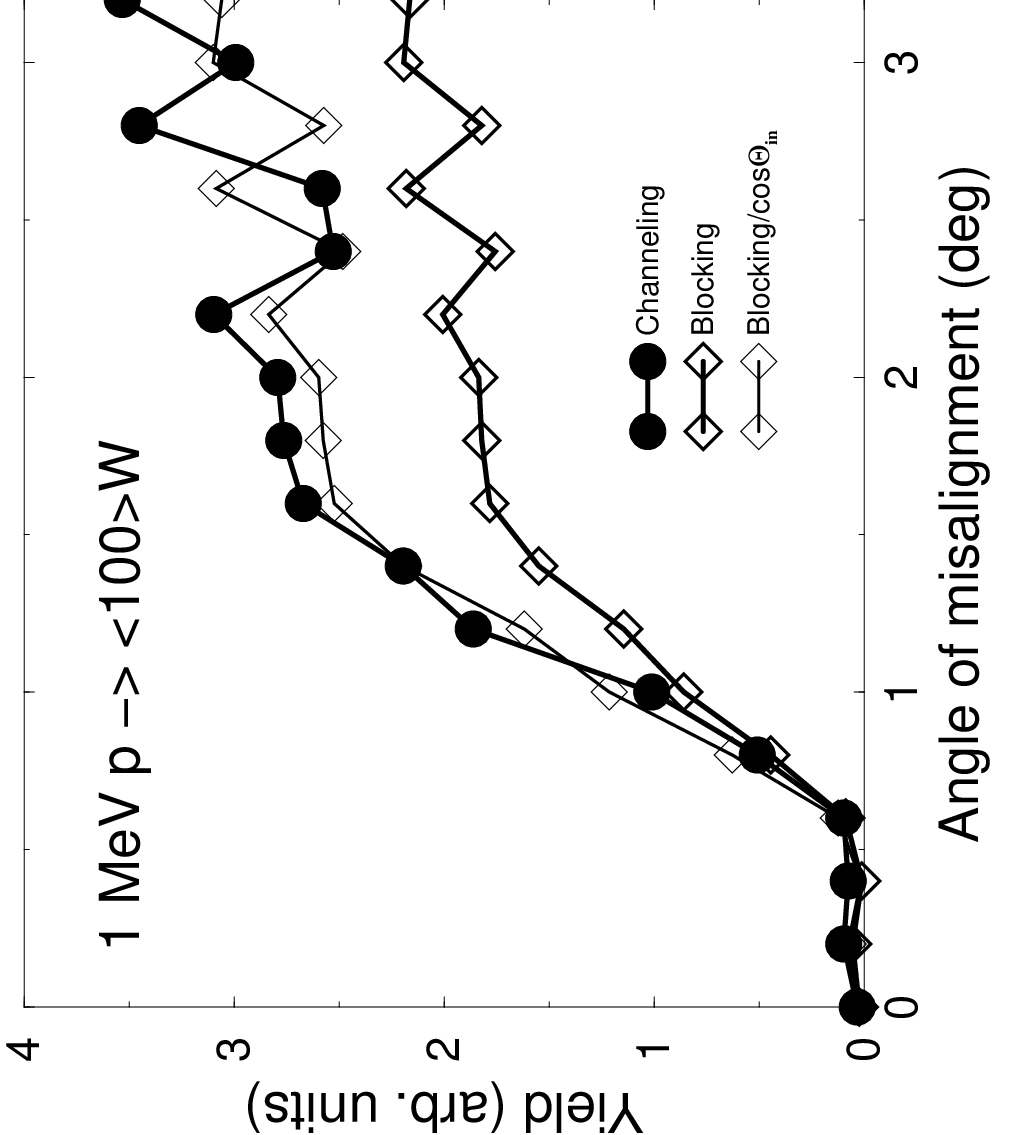}
\caption{\label{Fig8}
{\small
Channeling and blocking dips calculated for the case of irradiation with
1~MeV protons of a tungsten crystal along the $<$100$>$ axial direction
with detection at a 135$^\circ$ scattering angle and in the inverse geometry.
The yield is calculated in the energy window corresponding to scattering
from a depth range between 2500 and 3500~\AA. The blocking dip normalized to
the thickness of the layer along the beam direction is also shown.}
}
\end{figure}

As a last example we reproduce in our simulations an experiment~\cite{Bogh}
performed in the early times of channeling studies with the special aim to
test the Lindhard time-reversibility rule~\cite{Lindhard}. Here a proton
beam of high energy (1~MeV) is incident on a W crystal along the
$<$100$>$ direction (close to the surface normal) and the protons scattered
by 135$^\circ$ into a random direction of the crystal lattice are detected.
In the inverse situation the beam and detection directions were 
interchanged. The angular spread in the beam and the aperture of the 
detector were approximately equal, $0.1^\circ$, and the yield was 
measured within an energy window corresponding to a layer of 1000~{\AA } 
thickness at a depth of 3000~\AA. The yield was measured as a function 
of the beam misorientation in the first case and as a function of the 
detection angle relative to the $<$100$>$ channel in the second.
The simulated channeling and blocking wells are shown in Fig.~\ref{Fig8}. 
Their widths are similar and coincide well with the experimental results.
To achieve also an agreement between the absolute values, we had to
normalize the yield to the path length of the incoming ions inside
the considered layer (by multiplying the yield by $\cos\Theta_{in}$ in 
the second case, where $\Theta_{in}$ is the angle of beam incidence 
relative to the surface normal). In general, the time-reversibility 
is confirmed by the simulation similarly as in the referred experiment.

\section{Discussion}

In this section we discuss the advantages of the proposed approach in
comparison with those used currently in simulations of swift ion-solid
interaction. First, the shower approach is equivalent in accuracy
with the direct simulation as performed by the code MARLOWE.
The difference lies only in the much higher efficiency (lower level
of statistical fluctuations) as illustrated in Sect.~II and III.
This is in fact of primary importance since, in many situations,
extensive analysis of experimental data becomes feasible. Furthermore,
for medium and higher ion energies a proper simulation of multiple and
plural scattering is not possible at all by other methods.
For medium-energy ion scattering from amorphous samples, the accelerated
version of the code TRIM~\cite{Biersack} seems competitive but this is
achieved at the cost of additional approximations.

Barrett's approach~\cite{Barrett}, followed by later developments like
the programs FLUX~\cite{FLUX} and UPIC~\cite{UPIC}, is based on the
single-scattering model and has consequently its region of applicability
restricted to the cases where multiple and plural scattering can be
neglected. In principle, the calculation of the close-encounter
probabilities in this approach bears some similarity with our
procedure of shower generation. In this way, the picture of collisions
with small impact parameters is well reproduced. However, only one
of the two segments of the trajectory is described realistically
while the other is approximated by a straight line.

In fact, the only previously proposed method demonstrating, for the case of
low-energy ions, an adequate and, simultaneously, efficient treatment 
is the program MATCH~\cite{MATCH}. The time-reversing procedure used in this 
program is closely related with the shower approach and it is rather instructive 
to compare both strategies in detail. As found in simulations of LEIS (of the 
type shown in Fig.2) the use of the reversing approach is well competitive 
compared with the shower approach. It is not clear, however, if this is also 
the case in other conditions. In the rest of this section, we perform an analysis 
aiming to clarify whether there are some important differences in the basis of two
approaches, allowing in some cases to choose one of them as more convenient.
Note in advance that, in practical application, one would most probably prefer
the shower approach since, in its implementation, the reversing approach is 
much more cumbersome.

To clarify the above question, let us look closer at the basis of the reversing
approach. In the case of pure potential scattering, the time-reversibility
of ion motion suggests that the probability for an ion of the beam to
reach the detector and the probability of the inverse scattering are directly
related. The yield of scattering is simply proportional to the phase volume
of the detector, i.e., its acceptance. This property is a simple consequence 
of Liouville's theorem: the scattering yield is obtained as the overlap of 
the flux of all scattered ions with the detector acceptance and the volume 
of the region of overlap is invariant with respect to the time-translation. 
In the simulation, this means that there is no preference for the 
time-reversed mode compared to the direct reproduction of scattering; the 
two methods require comparable efforts. 
The presence of energy losses does not affect this conclusion. The reversing 
approach for simulation proposed in Ref.~[\onlinecite{VEGAS}] assumes 
a convolution of the beam profile with that of the detector at an 
intermediate position when ions reach a certain lattice site.

The procedure is illustrated in Fig.9. We consider the scattering of a single ion 
of the beam on a ``frozen" lattice additionally averaged over thermal displacements 
of one particular atom. In fact, the whole procedure of the simulation consists 
in the solution of such ``elementary" problems (see the first paragraph of Sect.~2). 
The region $\Phi_d^\prime$ shows schematically the detection profile shifted 
backwards in time to the considered site.
This region of phase space should be considered as actually 5-dimensional
(the longitudinal coordinate is not relevant). The possible states of
motion of the ion after collision with the atom of the considered site 
are distributed over a 3-dimensional hypersurface. The dimension is determined
by the dimension of the vector ${\bf R}$ of atom displacements. This 
lower-dimensional hypersurface is shown in Fig.9 schematically as the thick
solid curve. The probability of the interesting event of scattering can be
determined by the convolution of the distribution $\delta\Phi_b^\prime$
of ion states with $\Phi_d^\prime$.
\begin{figure}
\includegraphics*[width=70pt, height=200pt, angle=-90]{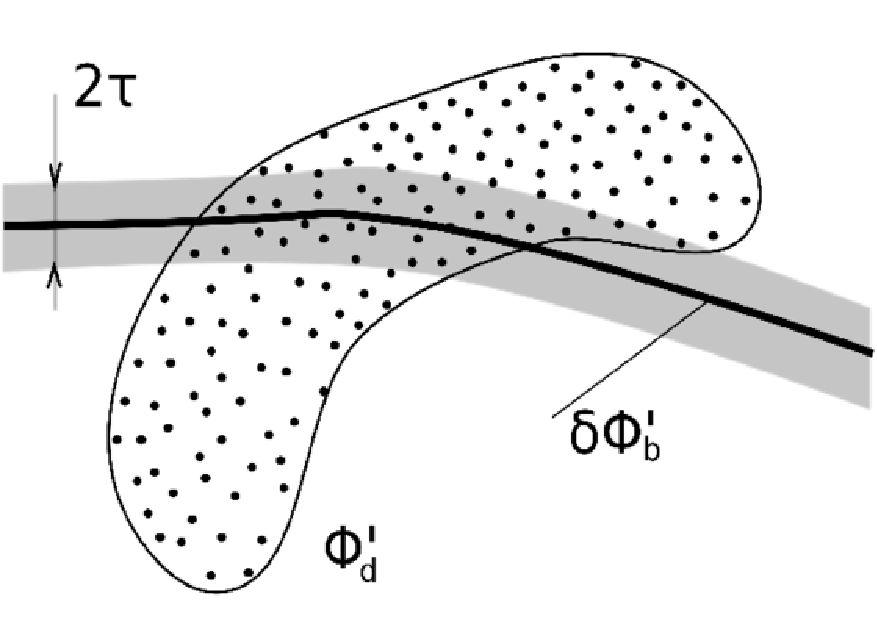}
\caption{\label{Fig9}
{\small
Schematic illustration of the procedure of flux convolution used in the
reversing approach (see the text for explanation of details).}
}
\end{figure}

In the simulation we represent the shifted detection profile 
$\Phi_d^\prime$ by a set of time-reversed trajectories ending up
in the detector (shown in Fig.9 by dots). Then, selecting those trajectories
which can be connected with the trajectory of the incoming ion, we can 
reconstruct specific examples of whole trajectories of detected ions. 
A connection takes place when some point in Fig.9 lies on the hypersurface 
$\delta\Phi_b^\prime$. The displacement of the atom ${\bf R}$
in the interesting collision is determined. 
It is clear, however, that exact connections are not probable. Thus, to
obtain sufficient statistics in the simulation we have to introduce a certain
tolerance for the connections. In Fig.9 this is illustrated as the shadowed
area around the hypersurface $\delta\Phi_b^\prime$: the points, states
of outgoing ions, which fall within this region are associated with possible
connections. 
Hence this approach assumes from the outset a certain approximation.
It is assumed in fact that the flux of ions before the collision is so
smooth that its variations within the tolerance region can be neglected.
Clearly, taking a weaker criterion for tolerance will increase the 
efficiency of the simulation. However, in the general case it is difficult
to estimate a priori how smooth the flux at a given site will be and this 
means that, strictly speaking, the resulting accuracy must be checked 
in each case by repeating the calculations with a tighter tolerance.

In the calculation of the scattering yield, the contributions of the found 
connections of trajectories are not equal but depend on the probability for 
the atom to be displaced to the required position ${\bf R}$ given by the 
distribution density $D({\bf R})$.
As derived in the Appendix, these contributions (the weights for the 
connections) are determined by the expression 
\begin{equation}\label{WWWW}
w=\frac{S_0\Delta \Omega\Delta E}{N_{out}}\cdot
\frac 1{\tau_y\tau_E\sin\Theta}
\frac {d \sigma}{d \Omega}
D({\bf R}).
\end{equation}
Here $\Theta$ is the angle of deflection in the ion-atom collision and
${d \sigma}/{d \Omega}$ is the differential cross section for ion 
scattering from the atom.
The tolerance for trajectory connection enters here through two parameters:
$\tau_y$, the distance between trajectories along the direction perpendicular
to the scattering plane and $\tau_E$, the discrepancy between energies.
The first fractional factor in the previous expression includes the 
dependence on all other parameters and ensures the correct absolute 
value of the yield.
These parameters are: the area of the unit cell of the crystal surface $S_0$,
the solid angle corresponding to the detector aperture $\Delta \Omega$,
the range of energies of the detected ions $\Delta E$ (the simulation 
gives the yield integrated over this range), and 
the number of calculated outgoing trajectories $N_{out}$.

Eq.(\ref{WWWW}) has a simple meaning. The first fractional factor in the
right-hand side is the phase volume per outgoing trajectory as it 
is determined by
the detector acceptance. The rest in the right-hand side is the density
of flux of the ion after the collision averaged over the distribution of 
atom thermal displacements ${\bf R}$. With non-zero tolerances $\tau_y$ and
$\tau_E$, the flux is distributed within a layer of finite ``thickness" around 
the original 3D hypersurface. 

One can readily see that this procedure also uses a special sampling of
atomic thermal displacements. Indeed, only those ${\bf R}$ are selected 
here which, with full certainty, result in interesting collisions. 
In this respect, the shower approach seems to be less efficient
because by far not every sampling of ${\bf R}$ results in a useful outcome.
However, one should additionally consider the following two circumstances.  
First, to shift the detection profile to the considered site we have to
calculate a sufficiently large number of outgoing trajectories and the
computational cost for this could be comparable with that in the direct
simulation. The second disadvantage in the reversing approach is the
possibility of strong fluctuations of the values of the weight 
[Eq.~(\ref{WWWW})],
first of all due to the strongly varying $D({\bf R})$. The relatively rare
events of plural scattering where all except one of the close collisions
are treated in the ordinary way can also result in exceedingly large
weights $w$. In such cases, additionally, the cross section $d\sigma/d\Omega$
can have large values.  It is easy to verify that the same variations of 
the weights would be found in the the direct simulation if one would
use an inconvenient uniform distribution instead of the natural (gaussian) 
distribution of ${\bf R}$. Such coincidence is not accidental: 
when pairs of trajectories are connected the corresponding displacement 
of the atom ${\bf R}$ is determined only by the kinematics of the binary 
collision and does not depend on anything else.

The yield of collisions with the atom, determined in this way, must be 
additionally averaged over the distribution of displacements of all
other atoms. This means that one needs to repeat the procedure described above 
for a sufficiently large number of randomly chosen configurations of the 
displacements. Furthermore, one should be careful when the total yield of 
scattering is determined as the sum of the contributions of different atoms. 
It is easy to understand that considering all possible connections of 
trajectories will inevitably result in repeated account of the same 
histories of motion. In fact the same problem is treated in the shower 
approach when we account for ``beam degradation". An analogous scheme could be 
applied also here though this would result in an exceedingly cumbersome
computational algorithm. The problem is partiallly eliminated in the algorithm
of the program MATCH where only close collisions are treated by trajectory 
connection. This approach works well when each ion experiences only one 
close collision but could fail in the treatment of plural scattering. 

The use of this approach turns out to be advantageous in two cases important 
for applications: scattering of medium-energy light ions, MEIS,
and low-energy heavy ion scattering, LEIS.
In both cases, simplified descriptions of ion-solid interaction
are possible without significant loss in accuracy. Both simplified
versions of the reversing approach are sufficiently efficient and, in the 
absence of other alternatives, they are is widely used in the analysis of 
experimental data. In MEIS (with a predominant use of 
beams of hydrogen ions) the picture
of single close collision is well adequate. As a first approximation, the 
fluctuations of the angle of scattering in the main collision can
be neglected. Also, the energy of ions is considered as directly related
to the depth of scattering. As a result, the description of ion fluxes is
reduced to the form of dependence on the transverse coordinates only.
The convolution of such fluxes, taking additionally into account the
probabilities of atom displacements and the scattering cross sections, is
a well tractable problem. This is the basis of the widely used program 
VEGAS. A more general approach is applied in the program SILISH
\cite{SILISH}. In this case the trajectories are connected accurately
neglecting only the energy relation. The results in
Ref.~[\onlinecite{SILISH}] show that such a minimal simplification turns out 
to be sufficient for the simulation to become possible, at least for 
the description of scattering from one monolayer.

The reversing approach is suitable to treat also LEIS of sufficiently 
heavy ions (the case of strong interaction). One example of results for such 
conditions is shown in Fig.~2. A favourable circumstance here is that 
only scattering from a few atomic layers is important. In addition, the large 
scattering cross section implies that the scattering on atoms results in 
strongly dispersed fluxes and, therefore, the result of their convolution 
is not very sensitive to the specific details of the flux distribution.
The program MATCH was developed to simulate scattering with such conditions.

Compared to the reversing approach, the simulation algorithm proposed
in the present paper treats scattering exactly within the binary 
collision model. At the same time, it demonstrates an unprecedented 
efficiency. It is also less cumbersome, the only difference with the 
direct method lies in the way the atom displacements are sampled. This means, 
in particular, that one can easily incorporate, if necessary, any additional 
features of the binary collisions like energy loss straggling or charge 
exchange.  Additionally, the shower approach is capable to reproduce 
in one run the energy spectrum of scattered ions (or recoiled atoms)
and also their angular distribution in a wide range. As demonstrated by 
the examples presented in Sect.~3, these two features make the method 
exceptionally powerful. Compared to this, the reversing approach is 
dramatically non-efficient. In fact, to obtain such results one has to 
repeat the calculations for each bin in the energy spectrum
(with a width $\Delta E$) and for each bin $\Delta \Omega$ of the
2D angular coordinates.

To finish this section we make some remarks intended to clarify possible
consequences of the reversibility rule for the interpretation of
experimental results. It is seen in Eq.(\ref{FFFF}) that, under the
assumption of a pure potential motion of the ions (the Jacobians
$J_{in}=J_{out}=1$), the yield of scattering from one lattice site is 
symmetric under exchange of the beam and detector directions 
provided that the flux in the beam is also uniform and the respective 
phase-space volumes are equal. The latter condition is less relevant 
because the difference can be simply accounted for by a factor in the 
yield. Therefore, we can conclude that the yield of scattering from 
one atom for a given beam-detector configuration is determined if it 
is known for the inverse situation. However, in the measurements of the 
yield in a certain energy window, as ordinarily made, the effective number 
of contributing atoms can be different. This fact was taken into account 
in the transformation of the data shown in Fig.~\ref{Fig8}.

Rigorously speaking, the reversibility rule is justified only when the 
picture of potential motion of the ions is assumed (pure potential 
scattering on infinitely heavy ions). In a real experiment this rule can 
be violated due to the recoiling of the atoms and due to the manifestation 
of their internal degrees of freedom, energy losses and multiple scattering 
on electrons. But in the performed simulation, these features were not 
considered and, in fact, this simulation is nothing more than a test of 
sensitivity to the round-off errors unavoidable in numerical calculations.
In principle, it is not obvious from the outset that the trajectories of 
ions and, consequently, the final results are stable with respect to these 
errors. Thus, any attempts to study the physical effects capable to lead 
to a violation of the reversibility rule should be preceded by a simulation 
as that performed here.

\section{Conclusions}

The shower approach proposed in this paper solves effectively the main
problem of simulation of ion scattering from solids within the binary
collision model, which is the elimination of the violent statistical
fluctuations in the Monte-Carlo sum. This is achieved by specific 
improvements of the direct simulation approach: the use of the strategies 
of importance- and stratified sampling. As a result, the computer power 
required for simulation is reduced by several orders of magnitude. 
This is, in fact, a decisive advantage allowing to address simulation problems
which cannot be treated with other methods.
As examples, we can mention the plural scattering of 
medium-energy ions and the simulations of 2D angular distributions. 
As discussed in Sect.~IV, our method avoids also many shortcomings 
inherent to alternative approaches. It is argued in particular that, 
in fact for the first time, this
method allows a reliable treatment of the rare events of plural
scattering. Such possibility is specially important because the plural
scattering is also not amenable to theoretical treatment.

We performed a detailed analysis of the approach based on the convolution of
fluxes of incoming and outgoing ions, as performed by the programs VEGAS and
MATCH. The latter program offers an alternative for an exact treatment
of the binary collision model, including multiple and plural scattering.
However, as follows from the arguments presented in Sec.4, the main illness
of the direct simulation, violent fluctuations in the accumulated statistics,
is inherited by this method. In general, the proposed shower approach
represents an effective replacement of widely used algorithms of
simulation providing qualitatively new possibilities for the analysis of
experimental results.

Simulations with the code TRIC can be performed for large classes of
crystal structures and provide a detailed picture of scattering or 
recoil yields in the form of energy and angular distributions. 
All these qualities are promising for a wide use of the developed 
computer code both in basic research and in the analysis of materials. 
In particular, this provides the possibility to efficiently compare 
measurements with simulations made for many trial structures allowing 
in this way precise structural analysis. Currently, the
alternative for analysis of MEIS results is the program VEGAS. It has, 
however, many restrictions in its application. The level of approximations 
used does not ensure a sufficient accuracy of the simulation for ions other 
than the lightest H and He. Also, this program does not provide energy 
spectra of scattered ions, an experimental result containg a large amount
of information. In addition, it is very hard to account with
this program for an intimate feature of lattice dynamics, the correlations
in thermal vibrations, to which data like those shown in Fig.6 can be
sensitive. VEGAS cannot help at all in the analysis of complex data 
measured with the modern technique, 3D-MEIS \cite{3DMEIS}, where
energy and angular distributions are simultaneously measured. In contrast, 
the shower approach is free of such limitations.

This paper shows several examples of the use of the program although,
of course, it is not possible to cover all potential applications
(e.g. simulations of the sputtering or total reflection yields, 
possible in this approach, are also interesting applications of the code 
TRIC). The illustration examples in Sect.~III are chosen to demonstrate the
capabilities to solve specific problems and to show the accuracy of
this method in comparison with others. In particular, we show the 
capability to simulate the interaction of different ions of low and medium 
energies with solid matter including complex structures, to calculate the 
yield of scattered ions and recoils and to reproduce their energy spectra 
and angular distributions. In addition, we address the interpretation of 
the time reversibility rule and provide additional insight into the origin 
of the surface peak.


\appendix
\section*{Appendix}
\setcounter{section}{1}

In this appendix, the expression Eq.~(\ref{WWWW}) for the weights assigned to
trajectory crossings in the MATCH approach is derived.

Let $\omega_1$ represent the phase variables of an ion impinging on the 
sample when it reaches the volume of the considered lattice site. The 
probability that, in the course of its further motion, the ion will end 
up in the detector is obtained as
\begin{equation}\label{AAAA}
P_i(\omega_1)=\int d\omega_2\frac {dP_{sc}}{d\omega_2}P_{out}.
\end{equation}
The integration is performed over the phase variables of the ion
after the collision $\omega_2$ and $({dP_{sc}}/{d\omega_2})d\omega_2$ is 
the probability of scattering into one of the states within $d\omega_2$; this
probability is related to the probability that the atom is located at the 
appropriate position. Finally, $P_{out}(\omega_2)$ is the probability that, 
in the course of its further motion, the ion in a state within $d\omega_2$ 
will leave the crystal with an energy and in a direction within the detector
acceptance.

The probability $d P_{sc}/d \omega_2$ is explicitly determined in the
case that both the scattering angle $\Theta({\bf b})$ and the energy
loss $\Delta E({\bf b})$ in the binary ion-atom collision are uniquely
determined by the impact parameter ${\bf b}$.
To derive the corresponding expression we describe the states $\omega_1$
and $\omega_2$ in a local coordinate system where the $z$ axis is aligned with
the ion velocity before the collision. As a result,
\begin{equation}\label{CCCC}
\frac {d P_{sc}}{d \omega_2}=\delta(y_2-y_1)\delta(E_2-E_1+\Delta E(b))\left|\frac{\partial^3{\bf R}}{\partial x_2 \partial^2 {\bf n}_2}\right|D({\bf R}),
\end{equation}
where $x$ and $y$ are the coordinates in the scattering plane and in the 
perpendicular direction, respectively. The first $\delta$-function 
satisfies the condition that the two trajectories must intersect and the 
second takes into account the relation between the energies before and 
after the collision. In the case of potential scattering, the state of 
ion motion after the collision (given by the coordinate in the scattering 
plane $x_2$ and the velocity direction ${\bf n}_2$) is uniquely determined 
by the atom position ${\bf R}$. The Jacobian of this relation appearing 
in (\ref{CCCC}) is expressed as

\begin{equation}\label{JJJJ}
\left|\frac{\partial^3{\bf R}}{\partial x_2 \partial^2 {\bf n}_2}\right|=\frac b{\sin^2\Theta}\left|\frac{d b}{d \Theta}\right|=\frac 1{\sin\Theta}\frac {d \sigma}{d \Omega},
\end{equation}
where $b$ is the impact parameter corresponding to the scattering angle
$\Theta$ and $d \sigma/d \Omega$ is the scattering cross section. Finally,
$D({\bf R})$ in (\ref{CCCC}) is the density of distribution of thermal
displacements of the target atom. Note that ${\bf R}$, the atom position
which results in the considered collision, is a function of $\omega_1$ and
$\omega_2$.

To confirm the validity of Eqs.(\ref{CCCC}) and (\ref{JJJJ}), let us 
calculate the angular dependence of the probability of scattering on a 
single atom (the variation due to the uncertainty of the atom position 
${\bf R}$). First, we integrate both sides of Eq.(\ref{CCCC}) over $y_2$ 
and $E_2$; this cancels the two $\delta$-functions. The coordinate $x_2$ 
is related to the coordinate $z_c$ of the point of crossing of the 
scattering asymptotes in such a way that
$dx_2=\sin \Theta \,dz_c$.
In addition, we take into account that, for a given scattering angle,
$z_c$ is directly related to the coordinate $z$ of the atom: $dz_c=dz$.
As a result we arrive at the familiar expression:
\begin{equation}\label{SSSS}
\frac {d^3 P_{sc}}{dz d^2{\bf n}_2}=
\frac {d \sigma}{d \Omega}D({\bf R})
\end{equation}
which expresses nothing but the concept of differential cross section
${d \sigma}/{d \Omega}$.

The contribution of scattering from the considered atom to the yield at 
the detector is obtained as an integral of the probability (\ref{AAAA}) 
multiplied by the flux $\Phi_{in}(\omega_1)$ of ions of the beam reaching 
this lattice site:
\begin{equation}\label{IIII}
Y_i=\int d\omega_1 \Phi_{in}(\omega_1)P_i(\omega_1).
\end{equation}
Furthermore, the variables $\omega_1$ and $\omega_2$ are uniquely related to
$\omega_{in}$ and $\omega_{out}$ respectively, the parameters of motion
of the ion when it enters (exits) the sample volume.
It is useful to refer directly to the latter parameters and, for this purpose,
we replace the integrations over $\omega_1$ and $\omega_2$ in (\ref{AAAA})
and (\ref{IIII}) by integrations over $\omega_{in}$ and $\omega_{out}$,
respectively.
In this case the Jacobian $J_2=|d\omega_2/d\omega_{out}|$ must be included 
in the integrand, while the flux $\Phi_{in}(\omega_1)$ is replaced by the 
flux of ions in the beam $\Phi_{b}(\omega_{in})$ multiplied by the Jacobian
$J_1=|d\omega_1/d\omega_{in}|$. Note that the relation $\omega_1(\omega_{in})$ 
cannot be defined for all $\omega_{in}$ (some initial conditions $\omega_{in}$ 
result in trajectories which never pass through the vicinity of the 
considered lattice site). The same is possible for the pair of variables
$\omega_2$ and $\omega_{out}$. To account for this, we assume the value
of the respective Jacobian in such cases to be zero. Now, combining the 
above equations, we arrive at the expression:
\begin{align}
Y_i&=\int d\omega_{out} J_{out}P_{out}
\int d\omega_{in} J_{in}\Phi_b\cdot \notag \\
&\cdot\delta(y_2-y_1)\delta(E_2-E_1+\Delta E(b))
\frac 1{\sin \Theta}
\frac {d \sigma}{d \Omega}
D({\bf R}) .\label{FFFF}
\end{align}
As a function of $\omega_{out}$, the probability $P_{out}$ is easily
evaluated: $P_{out}=1$ if $\omega_{out}$ is within the region of 
acceptance of the detector, otherwise $P_{out}=0$.

One can interpret the integration over $\omega_{out}$ in (\ref{FFFF}) as
a projection of the outgoing flux on the set $|d>$ of states determined 
by the detector acceptance; the ``density of states" is here uniform.
Denoting also the ``state of beam" (the distribution over $\omega_{in}$ in 
the beam) as $|b>$, we can formally write Eq.(\ref{FFFF}) as
\begin{equation}\label{YYYY}
Y_i=<d|{\cal T}_2{\cal S}_i{\cal T}_1|b>,
\end{equation}
where the operators of transformation of the fluxes ${\cal T}_1$ and 
${\cal T}_2$ are represented as the Jacobians $J_1$ and $J_2$, respectively
and the rest of the integrand in Eq.(\ref{FFFF}) represents the operator of
scattering on the atom ${\cal S}_i$. This interpretation shows how the 
integral Eq.(\ref{FFFF}) can be evaluated using the Monte-Carlo method. 
The distribution over $\omega_{in}$ for the impinging ions must be taken 
according to the beam profile. The transformation ${\cal T}_1|b>$ is obtained 
by calculating the trajectories of the incoming ions. The detection profile 
$|d>$ must be taken as a uniform distribution within the region of phase space
restricted by the unit cell at the sample surface $S_0$, by the solid angle
of detector acceptance $\Delta\Omega$ and by the considered energy
range $\Delta E$. The integral over $\omega_{out}$ is associated then
with a Monte-Carlo sum:
\begin{equation}\label{S1S1}
\int d\omega_{out}\Rightarrow \frac{S_0\Delta \Omega\Delta E}{N_{out}}
\sum_{k=1}^{N_{out}},
\end{equation}
where $N_{out}$ is the number of considered outgoing trajectories.
The set of outgoing trajectories calculated in time-reversed mode represents
the action of the operator inverse to ${\cal T}_2$. The fluxes of ingoing and
outgoing ions are convoluted by the matching of crossing trajectories and the
terms of the Monte-Carlo sum (\ref{S1S1}) are determined as the values of the
integrand in Eq.(\ref{FFFF}) To avoid the problems due to the presence of
$\delta$-functions in the integrand, they are replaced by normalized pulse
functions of finite width:
\begin{equation}\label{PPPP}
\delta(t)\Rightarrow \Pi(t)=\frac 1\tau\left\{
\begin{array}{rl}
1 & \rm{if} \; |t/\tau|<1/2\\
0 & \rm{if} \; |t/\tau|>1/2
\end{array} \right.
\end{equation}
This results immediately in the expression Eq.(\ref{WWWW}) of Sec.4
determining the terms $w$ of the sum (\ref{S1S1}), the weights of
crossing of trajectories. The tolerances, $\tau_y$ and $\tau_E$,
must be chosen to provide sufficient accuracy of the simulation results.

Note that in Ref.~[\onlinecite{MATCH}], where this approach was
proposed, the weight $w$ was simply taken as the product of the
atomic density $D({\bf R})$ with the cross section $d\sigma/d\Omega$.
The additional $\sin \Theta$ in the denominator of Eq.~(\ref{WWWW})
accounts for the fact that the density of crossings of two set
of parallel trajectories is inversely proportional to $\sin \Theta$,
merely a simple geometrical feature. Additionally, the first fractional
factor provides the correct normalization of the fluxes.

\acknowledgments V.A. Khodyrev gratefully acknowledges the hospitality provided by the
CMAM during several research stays. The authors want to thank A. Guirao and
J. \'Alvarez from CMAM for technical help with computer resources. This
work was financed by projects MAT2005-03011 and FIS2008-01431; J.E. Prieto
and A. Rivera also acknowledge support by the programs ``Ram\'on y Cajal"
and ``J.~de la Cierva" of the Spanish MEC, respectively.


\end{document}